\documentclass[%
reprint,
 superscriptaddress,
nofootinbib,
 amsmath,amssymb,
 aps,
 prd,
floatfix,
]{revtex4-2}

\usepackage{graphicx,amssymb,amsmath,amsthm,amsfonts,epsfig,epsf}
\usepackage[linktocpage]{hyperref}
\usepackage[usenames]{color}
\usepackage{epstopdf}
\usepackage{bm}
\usepackage{dcolumn}
\usepackage[latin1]{inputenc}
\usepackage{latexsym}
\usepackage{rotating}
\usepackage{hyperref}
\usepackage{color}
\usepackage{longtable}

\usepackage{enumerate}
\usepackage{tensor,multirow}
\usepackage{url}
\usepackage{tikz,tikz-3dplot}
\usepackage{ulem}

\usepackage{slashed,orcidlink}

\newcommand{\nn}{\nonumber}

\usepackage[scr=boondox]{mathalfa}

\newcommand{\cm}{\mathscr{m}}

\newcommand{\cb}{\mathscr{b}}
\newcommand{\ca}{\mathscr{a}}
\newcommand{\cc}{\mathscr{c}}

\DeclareFontFamily{U}{min}{}
\DeclareFontShape{U}{min}{m}{n}{<-> udmj30}{}

\newcommand{\GSSI}{Gran Sasso Science Institute (GSSI), I-67100 L'Aquila, Italy}
\newcommand{\GranSasso}{INFN, Laboratori Nazionali del Gran Sasso, I-67100 Assergi, Italy}
\newcommand{\milan}{\affiliation{Dipartimento di Fisica ``G. Occhialini'', 
Universit\'a degli Studi di Milano-Bicocca, Piazza della Scienza 3, 20126 Milano, Italy}}
\newcommand{\infn}{\affiliation{INFN, Sezione di Milano-Bicocca, 
Piazza della Scienza 3, 20126 Milano, Italy}}

\begin{document}

\preprint{ET-0312A-23}

\title{Relevance of Precession for Tests of the Black Hole No Hair Theorems}

\author{Nicholas Loutrel\orcidlink{0000-0002-1597-3281}}
 \email{nicholas.loutrel@unimib.it}
 \milan
 \infn
 \affiliation{Dipartimento di Fisica, ``Sapienza'' Universit\`a di Roma \& Sezione INFN Roma1, Piazzale Aldo Moro 5, 00185, Roma, Italy}
 
\author{Richard Brito\orcidlink{0000-0002-7807-3053}}
\affiliation{CENTRA, Departamento de F\'{\i}sica, Instituto Superior T\'ecnico -- IST, Universidade de Lisboa -- UL, Avenida Rovisco Pais 1, 1049 Lisboa, Portugal}

\author{Andrea Maselli\orcidlink{0000-0001-8515-8525}}
\affiliation{\GSSI}
\affiliation{\GranSasso}

\author{Paolo Pani\orcidlink{0000-0003-4443-1761}}
\affiliation{Dipartimento di Fisica, ``Sapienza'' Universit\`a di Roma \& Sezione INFN Roma1, Piazzale Aldo Moro 5, 00185, Roma, Italy}

\date{\today}

\begin{abstract}
The multipole moments of black holes in general relativity obey certain consistency relations known as the no-hair theorems. The details of this multipolar structure are imprinted into the gravitational waves emitted by binary black holes, particularly if the binary is precessing. If black holes do not obey the vacuum field equations of general relativity, then the no-hair theorems may be broken, and the observed gravitational waves will be modified, thus providing an important test of the no-hair theorems. Recently, analytic solutions to the precession dynamics and inspiral waveforms were computed within the context of binaries possessing non-axisymmetric mass quadrupole moments, which are parametrized by a modulus $q_{\cm}$ and phase $a_{\cm}$ with $\cm = 1,2$ the azimuthal spherical harmonic number. Here, we use a Fisher analysis to study plausible constraints one may obtain on generic, non-axisymmetry quadrupole configurations using current and future ground-based detectors. For non-precessing binaries, we generically find that no meaningful constraints can be placed with current detectors on the non-axisymmetry parameters $(q_{\cm}, a_{\cm})$ due to the presence of strong degeneracies with other waveform parameters, while with next generation detectors, only weak constraints are possible. For precessing configurations, the exact value of the uncertainty is strongly dependent on the sky location, system orientation relative to the line of sight, and initial inclination angle of the orbital angular momentum. After averaging over these parameters, we find that with GWTC-3-like events, one should be able to plausibly constraint non-axisymmetric mass quadrupole deviations to $\Delta q_{\cm} \sim 10^{-2}$ for LIGO at design sensitivity, and $\Delta q_{\cm} \sim 10^{-4}$ for the same sources with Einstein Telescope and Cosmic Explorer.
\end{abstract}

\maketitle



\section{Introduction}
\label{sec:intro}

A fundamental prediction of black holes (BHs) as stationary vacuum solutions to the field equations of general relativity (GR) is the so-called no hair theorems~\cite{Hawking:1971vc,Hawking:1973uf,Carter:1971zc,Robinson:1975bv}. The theorems state that the object's $l\ge 2$ multipole moments\footnote{Here, $(l,m)$ refer to the spherical harmonic degree and order.} are only dependent on the mass monopole, spin dipole, and electric monopole charge, the latter of which is not expected to be of relevance to astrophysical BHs. The higher multipole structure of the BHs is spin-induced, i.e. the presence of spin angular momentum uniquely determines the higher order multipole moments of the object. As a result, all BHs in GR are axisymmetric, specifically, all multipole moments with $m\neq 0$ vanish.

However, the theorems are unique to BHs as described by the Kerr-Newman metric~\cite{KerrNewman}, the most general stationary vacuum solution of GR~\cite{Campanelli:2008dv}. In modified theories of gravity, it is well-known that the no-hair theorems can be violated (see, for example, ~\cite{Chrusciel:2012jk,Berti:2015itd}), and as a result the $l\ge 2$ multipole moments of BHs are modified from their GR values. In many of these scenarios, the BHs can possess non-trivial hair generated from additional scalar, vector, or tensor fields. As a result, the multipolar structure is no longer determined only by the BH mass and spin angular momentum, but also additional hair.

For non-vacuum configurations, both astrophysical and exotic, the no hair theorems generally do not hold. For astrophysical compact objects where the higher multipole moments are spin-induced, the objects still possess a spheroidal shape, but the precise values are equation of state dependent~\cite{Laarakkers:1997hb}. Furthermore, in exotic scenarios where the compact object is not composed of ordinary matter~\cite{Liebling:2012fv,Herdeiro:2020kvf,Raposo:2020yjy,Ildefonso:2023qty,Herdeiro:2023roz,Cunha:2022tvk,Sanchis-Gual:2021edp}, or is the result of low energy limit of a UV complete theory~\cite{Bena:2020see,Bena:2020uup,Bianchi:2020bxa,Bianchi:2020miz,Bah:2021jno,Bena:2022rna}, axisymmetry can be broken, leading to the presence of non-vanishing $m\neq 0$ multipole moments. 

The multipolar structure of compact objects has a direct impact on their observed gravitational wave (GW) emission. It is well known that, even at the Newtonian level, objects with rich multipolar structure will induce precession of the orbital plane when in a binary~\cite{PoissonWill:sec:3.4}. Relativistic Lens-Thirring and spin-spin interactions further induce so-called spin precession, which has been extensively studied in GR (see e.g. ~\cite{Chatziioannou:2016ezg,Chatziioannou:2017tdw,Kesden:2014sla,Racine:2008qv,Gerosa:2015hba,Gerosa:2015tea,Zhao:2017tro,Khan:2015jqa,Khan:2018fmp,Khan:2019kot}) and in modified theories of gravity~\cite{Loutrel:2018ydv,Brax:2021qqo,Alexander:2017jmt}. Precession necessarily induces modulation of the GW amplitude, phase, and frequency~\cite{Apostolatos:1994}, which are encoded in additional harmonics beyond the standard orbital harmonics of the waveform~\cite{Chatziioannou:2016ezg,Blanchet:2013haa}. Each orbital harmonic possesses a set of precessional ``overtones," specifically $m'$ harmonics that encode simple precession effects, and $n$ harmonics which encode nutation effects~\cite{Loutrel:2022xok}. The inclusion of these in waveform models has been shown to be critical in breaking parameter degeneracies in data analysis settings~\cite{Chatziioannou:2014coa}.

Detection, or non-detection, of non-axisymmetric mass quadrupole configurations in precessing binaries provides a smoking gun test of the no hair theorems. However, most studies of precessional effects are limited to spin-induced, spheroidal configurations~\cite{Chatziioannou:2017tdw,Gerosa:2015hba,Gerosa:2015tea,Kesden:2014sla}. Recently, precession dynamics for compact objects with generic mass quadrupole moments was considered in~\cite{Loutrel:2022ant} using the post-Newtonian (PN) approximation. At lowest PN order, the multipolar structure of the object enters the equations of motion at second PN (or 2PN) order through the monopole-quadrupole interaction. The non-axisymmetry of the compact objects are then encoded through the complex mass quadrupole coefficients $Q_{\cm}$, where $\cm = \pm 1, \pm2$. Since these coefficients are complex, they can be recast in terms of real-valued modulus $q_{\cm}$ and argument (or phase) $a_{\cm}$ parameters. For $|\cm|=1$, these parameters are referred to as \textit{axial} parameters, due to the object possessing parity-odd non-axisymmetry. Similarly, for $|\cm|=2$, these parameters are referred to as \textit{polar}, and the object will have parity-even non-axisymmetry. In~\cite{Loutrel:2022ant}, analytic solutions for the orbital, precession, and dissipative dynamics were found for $q_{\cm} < 1$ and arbitrary $a_{\cm}$ using a multiple scale analysis~\cite{Bender-msa}. Analytic waveforms were then computed using the stationary phase approximation (SPA)~\cite{Bender-spa} and shifted uniform asymptotics (SUA), which were originally used to develop PN waveforms for spin precessing binaries in GR~\cite{Klein:2014bua,Chatziioannou:2016ezg}. 

These waveforms served as a useful example for proposing an extension of the parameterized post-Einsteinian (ppE) formalism~\cite{Yunes:2009ke} for precessing binaries in~\cite{Loutrel:2022xok}. While the standard non-precessing ppE formalism only require two deviations, one for the waveform amplitude and one for the waveform phase, to capture the leading order beyond-GR effects, the analysis in~\cite{Loutrel:2022xok} found that significantly more may be necessary for a precessing extension of the ppE framework. To cleanly answer how many ppE parameters are needed for precessing binaries, one would need to perform an in-depth parameter estimation study for each beyond-GR scenario.

Here, we extend the previous studies of~\cite{Loutrel:2022ant,Loutrel:2022xok} to determine what are the ``correct" ppE parameters for testing the no-hair theorems with generic mass quadrupole effects, and what are the reasonable constraints one may be able to obtain with observations of GWs from precessing binaries with current and future ground-based detectors. To do so, we perform a Fisher analysis on binary BH~(BBH) signals whose masses are chosen to correspond to a subset of the GWTC-3 events~\cite{GWTC3} that have evidence of spin precession. While these events are consistent with spin precessing binaries as described by GR, we use them as benchmarks for the present study where only quadrupole-monopole precession~\footnote{Precession induced by the 2PN monopole-quadrupole interaction, where the orbital angular momentum and body axis are mis-aligned.} is considered.

Generically, we find that the uncertainty on the $(q_{\cm}, a_{\cm})$ parameters are degenerate with many of the other waveform parameters, which is due in part to the fact that all mass-quadrupole effects enter the waveform phase at 2PN order. To address this, we perform a Bessel decomposition of the waveform phase to include oscillatory corrections to the precession dynamics that were originally found in~\cite{Loutrel:2022ant} but neglected in~\cite{Loutrel:2022xok}. Doing so, we find that axial non-axisymmetric parameters $(q_{1}, a_{1})$ only appear in the $n=\pm1$ waveform nutation overtones, while polar non-axisymmetric parameters $(q_{2}, a_{2})$ only appear in the $n=\pm2$ nutation overtones. As a result, there is a clean splitting between axial and polar asymmetries, and ppE analyses can be carried out for each case separately. We find that, in both cases, the $m'=0$ overtone of the waveform is most important in a ppE-style analysis, and we identify the ppE parameters for axial and polar asymmetries, respectively.

Once the correct ppE parameters have been identified, we study the constraints one can place with current and future detectors. Generically, we find that meaningful constraints on the argument parameters $a_{\cm}$ cannot be obtained due to the uncertainty being dominated by the range of the prior. On the other hand, we find the modulus parameters $q_{\cm}$ can be reasonably constrained to $\sim 10^{-4}$ for BBH events with signal-to-noise ratio (SNR) one thousand, as expected for the Einstein Telescope (ET)~\cite{Freise:2008dk,Punturo:2010zz,Hild:2010id,Maggiore:2019uih, Kalogera:2021bya,Branchesi:2023mws} and Cosmic Explorer (CE)~\cite{LIGOScientific:2016wof,Essick:2017wyl,Reitze:2019iox,Kalogera:2021bya}, while to $\sim 10^{-2}$ for events with SNR one hundred, as expected in LIGO~\cite{aligo} at design sensitivity. We repeat the analysis with the same signals, but in non-precessing configurations. For sources in LIGO, the uncertainty on $q_{\cm}$ spans the range of the prior ($q_{\cm} < 1$), and no meaningful constraints can be obtained with non-precessing signal. For the same signals in CE and ET, the same uncertainties are typically $~0.55-1.0$ due to the increased in SNR from next-generation detectors, but still three orders of magnitude weaker than those obtained from precessing configurations. This highlights the importance of precession dynamics for future tests of GR.

The remainder of this paper is organized as follows. Sec.~\ref{sec:theory} provides the details of precessing ppE waveform model, our method for including oscillatory effects in the precession dynamics, and the details of our Fisher analysis. In Sec.~\ref{sec:results1}, we provide the results of the Fisher analysis, with Sec.~\ref{sec:deg} giving a detailed discussion of parameter degeneracies, Sec.~\ref{sec:ppE} providing the details of our investigation into the proper ppE parameters for non-axisymmetric mass quadrupoles, and Sec.~\ref{sec:bounds} finally studying constraints on axisymmetry in BBH events. The new ppE parameters for this scenario are given in Eqs.~\eqref{eq:bppE-ax}-\eqref{eq:appE-pol}, while the projected constraints on violation of the no hair theorems with GWTC-3-like events are given in Tables~\ref{tab_et}-\ref{tab_ligo}. In Sec.~\ref{sec:disc}, we conclude with future directions. Throughout this work, we use units where $G=c=1$.

\section{Theory \& Methodology}
\label{sec:theory}

Here, we introduce the waveforms used in our analysis, and provide a broad overview of the Fisher matrix method for obtaining uncertainties on the waveform parameters.

\subsection{Precessing ppE Waveforms for Non-axisymmetric Quadrupoles}
\label{sec:waveform}

Consider a binary comprised of two compact objects with masses $m_{1,2}$ and mass quadrupole moments $Q_{1,2}^{<ij>}$. The dynamics of the perturbed two body problem reduce to an effective one body description with an effective mass quadrupole moment~\cite{Loutrel:2022ant}
\begin{equation}
    \label{eq:Qeff}
    Q_{\rm eff}^{<ij>} = \eta_{1} Q_{2}^{<ij>} + \eta_{2} Q_{1}^{<ij>}\,.
\end{equation}
where $\eta_{A} = m_{A}/M$ with $M = m_{1} + m_{2}$ the total mass of the binary. Note that Eq.~\eqref{eq:Qeff} corrects a typo in the expression of $Q_{\rm eff}$ below Eq. (8) of~\cite{Loutrel:2022ant}. The effective quadrupole tensor admits a decomposition into $l=2$ spherical harmonics, with complex coefficients $Q_{\cm}$. The coefficients with $\cm=0$ corresponds to spheroidal deformations, while $\cm = \pm1$ and $\cm=\pm2$ correspond to axial and polar quadrupole deformations, respectively. The coefficients with $\cm \neq 0$ break the axisymmetry of the compact objects. From these, one can define the reduced parameters $\chi_{Q} = Q_{0}/M^{3} \eta$ and
\begin{equation}
    \label{eq:non-axi-defs}
    q_{|\cm|} e^{i \cm a_{|\cm|}} = \sqrt{\frac{2}{3}} \frac{Q_{\cm}}{Q_{0}}\,,
\end{equation}
with $q_{|\cm|}$ the modulus and $a_{|\cm|}$ the argument of the non-axisymmetric deformations\footnote{The $(q_{\cm}, a_{\cm})$ are derived from both $Q_{\cm}$ and $Q_{-\cm}$ coefficients, since $Q_{-|\cm|} = (-1)^{\cm} Q_{\cm}^{\dagger}$, where $\dagger$ corresponds to complex conjugation. As a result, the $\cm$ index on $(q_{\cm}, a_{\cm})$ only takes positive values, resulting in the absolute values in Eq.~\eqref{eq:non-axi-defs}. For the remainder of the paper, we drop the absolute value on $\cm$ in these parameters.}. 

Analytic, inspiral-only, frequency domain waveforms for binaries precessing due to generic quadrupole-monopole effects were derived in~\cite{Loutrel:2022ant}, but simplified into the ppE formalism in~\cite{Loutrel:2022xok}. For the analysis carried out herein, we use the latter, where the waveform polarizations take the form
\begin{align}
    \label{eq:ppE-wave}
    \tilde{h}_{+} - i \tilde{h}_{\times} &= \sum_{K} \tilde{h}_{K}^{\rm GR}(f) e^{i \cb_{\slashed{K}}^{\rm ppE} \times \left[U_{\rm ppE}(f)\right]^{b_{\slashed{K}}^{\rm ppE}}} 
    \nn \\
    &\times \left\{1 + \ca_{\slashed{K}}^{\rm ppE} \times \left[u_{k}^{\rm GR}(f) \right]^{a_{\slashed{K}}^{\rm ppE}} \right\}\,,
\end{align}
In the above expression, $\tilde{u} = (\pi M f)^{1/3}$ is the PN expansion parameter, $K = lmm'nk$ is a hyper-index\footnote{Note that here $m$ should not be confused with the harmonic number $\cm$ that appears in Eq.~\eqref{eq:non-axi-defs}, since they correspond to spherical harmonic decompositions of two separate quantities.}, and $\tilde{h}_{K}^{\rm GR}$ is given by
\begin{align}
    \tilde{h}_{K}^{\rm GR}(f) = {\cal{A}}_{K}^{\rm GR}(f) e^{i \Psi_{m}^{\rm GR}(f) + i \Phi_{mm'n}^{\rm P,GR}(f)} {_{-2}}Y_{lm'}(\theta_{N}, \phi_{N})
\end{align}
with $_{-2}Y_{lm}(\theta,\phi)$ spin-weight $s=-2$ spherical harmonics,  ${\cal{A}}_{K}^{\rm GR}$ given in Eq.~(101) in~\cite{Loutrel:2022xok}, and $\Psi_{m}^{\rm GR}$ the Fourier phase given in Eq.~(2) therein. The precession phase $\Phi_{mm'n}^{\rm P, GR} = m \epsilon(f) + m' \alpha(f) + n \psi(f)$ is given by Eq.~(10) therein, with replacement $u \rightarrow u_{k}^{\rm GR}(f)$, where $u_{k}(f)$ is the SUA function given by Eq.~(6) therein. The frequency dependent behavior of the precession angles $[\epsilon, \alpha, \psi]$ are provided in Sec.~IIC therein (also Fig.~1 in~\cite{Loutrel:2022xok}). The function $U_{\rm ppE}(f)$ is determined by which part of the phase (orbital or precessing) gives the lowest PN order deviation from GR, and is given explicitly in Eq.~(91) in~\cite{Loutrel:2022xok}.

The precessing ppE waveform in Eq.~\eqref{eq:ppE-wave} is \textit{polyphonic}, meaning that it contains multiple harmonics (or ``voices"), as opposed to the original non-precessing ppE formalism for quasi-circular binaries~\cite{Yunes:2009ke,Loutrel:2022xok}. Each harmonic index contained in the hyper index $K$ corresponds to physical effects, specifically
\begin{itemize}
    \item $l$: Polar harmonic number corresponding to the order of radiation reaction effects in PN theory (i.e. $l=2$ quadrupole radiation, $l=3$ octopole radiation, etc),
    \item $m$: Azimuthal harmonic number corresponding to harmonics of the carrier phase $\phi_{C}$, which includes contributions from the orbital phase $\phi_{\rm orb}$ and Thomas phase $\epsilon$. The azimuthal number is bounded by $|m| \le l$. 
    \item $m'$: Precession ``overtones" of each carrier harmonic $m$ due to the azimuthal motion of the orbital angular momentum, and characterized by the precession angle $\alpha$. The precession number is bounded by $|m'| \le l$.
    \item $n$: Nutation ``overtones" of each azimuthal harmonic $m$ due to the polar motion of the orbital angular momentum, and characterized by the nutation phase $\psi$ of the inclination angle $\beta = \beta(\psi)$. Generically, there is no bound on the nutation number $n$, which varies based on the specific precessing scenario.
\end{itemize}
The index $k$, on the other hand, does not correspond to new harmonic information, but arises as a result of the SUA re-summation procedure~\cite{Klein:2014bua,Chatziioannou:2016ezg,Chatziioannou:2017tdw}, and is bounded by $k_{\rm max}$, which is determined by waveform accuracy and computational efficiency considerations.  For the analysis carried out in our previous work and herein, $l=2$ since we work at leading PN order in the waveform amplitude, and $m=-2$ due to the intricacies of the SPA+SUA scheme used to derive Eq.~\eqref{eq:ppE-wave}. For more generic scenarios where higher waveforms harmonics are important, one does not need to be as restrictive, and can sum over multiple $(l,m)$ modes.

For the non-GR sector of the waveform in Eq.~\eqref{eq:ppE-wave}, $\slashed{K}$ is a subset of $K$, $[\cb_{\slashed{K}}^{\rm ppE}, \ca_{\slashed{K}}^{\rm ppE}]$ are ppE amplitude parameters, and $[b_{\slashed{K}}^{\rm ppE}, a_{\slashed{K}}^{\rm ppE}]$ are ppE exponent parameters. Since $\slashed{K}$ is a subset of the hyper-index $K$, only a subset of the total harmonics of the waveform contain ppE deformations. For the waveforms considered in~\cite{Loutrel:2022ant}, $l=2$ and $m=-2$, and thus the sum over $K$ reduces to a sum over the precession and nutation harmonics numbers $(m',n)$. By considering a simplified likelihood analysis, ~\cite{Loutrel:2022xok} proposed that five harmonics may need ppE deformations, with the ppE parameters specifically given in Eqs.~(118)-(123) \& (142)-(143) therein, and in Appendix~\ref{app:coeffs} herein for completeness. As part of this study, we determine whether all of these ppE corrections are needed to place constraints on the non-axisymmetry parameters $[q_{\cm}, a_{\cm}]$, as well as whether these are the ``correct" ppE parameters. 

With the waveform polarizations in Eq.~\eqref{eq:ppE-wave}, the detector response is given by
\begin{equation}
    h = F_{+}(\theta_{S}, \phi_{S}, \psi_{\rm pol}) h_{+} + F_{\times}(\theta_{S}, \phi_{S}, \psi_{\rm pol}) h_{\times}
\end{equation}
where $F_{+/\times}$ are the detector beam pattern functions given by~\cite{Apostolatos:1994,Stavridis:2009mb}
\begin{align}
    F_{+} &= \frac{1}{2} \left(1 + \cos^{2}\theta_{S}\right) \cos(2\phi_{S}) \cos(2\psi_{\rm pol}) 
    \nn \\
    &- \cos\theta_{S} \sin(2\phi_{S}) \sin(2\psi_{\rm pol})\,,
    \\
    F_{\times} &= \frac{1}{2} \left(1 + \cos^{2}\theta_{S}\right) \cos(2\phi_{S}) \sin(2\psi_{\rm pol}) 
    \nn \\
    &+ \cos\theta_{S} \sin(2\phi_{S}) \cos(2\psi_{\rm pol})\,,
\end{align}
with $(\theta_{S},\phi_{S})$ the sky location of the source and $\psi_{\rm pol}$ the time-dependent polarization angle,
\begin{equation}
    \label{eq:pol-angle}
    \psi_{\rm pol} = \tan^{-1} \left[\frac{(\hat{L} \cdot \hat{z}) - (\hat{L} \cdot \hat{N}) (\hat{z} \cdot \hat{N})}{\hat{N} \cdot (\hat{L} \times \hat{z})} \right]\,.
\end{equation}
In the above equation, $\hat{N}$ is the line of sight to the source/detector, $\hat{z}$ is the z-axis of the detector, and $\hat{L}$ is the orbital angular momentum, which evolves on the precession timescale according to Eq.~(28) in~\cite{Loutrel:2022ant}. When computing the detector response in the Fourier domain, the beam pattern functions must be pulled into the sum over the index $k$, since the polarization angle becomes dependent on the SUA function, i.e $\psi_{\rm pol} \rightarrow \psi_{\rm pol}[u_{k}^{\rm GR}(f)]$. 

The completed waveform (or rather detector response) is parameterized by eighteen parameters, specifically:
\begin{itemize}
    \item $M = m_{1} + m_{2}$: total mass of the binary
    \item $\eta = m_{1} m_{2}/M^{2}$: symmetric mass ratio
    \item $\chi_{Q}$: dimensionless spheroidal quadrupole parameter
    \item $q_{1,2}$: axial and polar non-axisymmetry modulus parameters
    \item $a_{1,2}$: axial and polar non-axisymmetry argument parameters
    \item $D_{L}$: luminosity distance to the source
    \item $(t_{c}, \phi_{c})$: time and orbital phase of coalescence
    \item $(\beta_{0}, \omega_{0})$: inclination angle and longitude of pericenter, parametrizing the initial orientation of the orbital angular momentum vector, relative to the fixed axis of precession
    \item $(\psi_{c}, \epsilon_{c})$: precession phases of coalescence
    \item $(\theta_{N}, \phi_{N})$: orientation of the direction of propagation $\hat{N}$ in the binary's frame
    \item $(\theta_{S}, \phi_{S})$: sky location of the source in the detector frame
\end{itemize}
Only four of these parameters, specifically $(q_{\cm}, a_{\cm})$, qualify as ``beyond-GR" (more specifically, beyond-vacuum-GR) parameters since they are zero for GR BHs. 

Before continuing, we note that precession in our analysis only refers to quadrupole-monopole precession of the orbital plane, and thus orbital angular momentum. Following~\cite{Loutrel:2022ant}, we neglect all effects induced by spin angular momentum of the component objects of the binary as a simplifying assumptions. We will only consider spin effects when determining reasonable values of $\chi_{Q}$ for BBHs in Sec.~\ref{sec:fisher}.

\subsection{Oscillatory Contributions \& Nutation Harmonics}
\label{sec:osc}

The ppE waveform of Eq.~\eqref{eq:ppE-wave} was derived in~\cite{Loutrel:2022xok}, where it was proposed that the beyond-GR correction to the waveform phase takes the form
\begin{equation}
    \label{eq:delta-Psi}
    \delta \Psi(f) = \cb \tilde{u}^{-1} + \cc_{(m',n)} \times \left[u_{k}^{\rm GR}(f)\right]^{-1}\,,
\end{equation}
where $(\cb, \cc)$ are given in Appendix~\ref{app:coeffs} and depend on the non-axisymmetry parameters $(q_{\cm}, a_{\cm})$, angles $(\beta_{0}, \omega_{0})$, and spheroidal quadrupole parameter $\chi_{Q}$. The first term comes from the SPA, while the second term arises due to the SUA re-summation, which is why it depends on the SUA function $u_{k}^{\rm GR}(f)$. Both of these terms enter at 2PN order, the latter in an effective sense since $u_{k}^{\rm GR}(f) \sim \tilde{u} + {\cal{O}}(\tilde{u}^{7/2})$. As a result, the non-axisymmetry parameters $(q_{\cm}, a_{\cm})$ are all degenerate with one another, as well as being degenerate with the angles $(\beta_{0},\omega_{0})$, and the parameter $\chi_{Q}$. Further,~\cite{Loutrel:2022xok} showed that there is a strong correlation between non-axisymmetry effects and both the sky location $(\theta_{S}, \phi_{S})$ and direction of propagation $(\theta_{N}, \phi_{N})$. These issues, the degeneracies and strong correlations, pose a challenge when performing parameter estimation, even in the simplified analysis carried out herein. This is especially true given that one of our goals is to identify the most important ppE effects in the waveform of Eq.~\eqref{eq:ppE-wave}, which is difficult to do when these issues are present. In this work we focus on the 
relevance of correlations among the intrinsic 
parameters of Eq.~\eqref{eq:delta-Psi}, 
averaging with respect to both sets of angles, 
the details of which are provided in Sec.~\ref{sec:avg}.
Thus, we are left with how to deal with the degeneracies present in Eq.~\eqref{eq:delta-Psi}.

The problem of the degeneracies in the non-axisymmetry parameters reduces to a more fundamental question. When performing null hypothesis tests of the no hair theorems, how does one distinguish between axial and polar non-axisymmetry breaking from Eq.~\eqref{eq:delta-Psi}? The answer is that one cannot in a data analysis scenario corresponding to how tests of GR are performed with current observations. Under the current methods, a bound would be placed upon the deviations to the 2PN coefficient of the waveform phase in GR. This is equivalent to a bound on the combination $\tilde{u} \times \delta\Psi(f)$, and one cannot disentangle the axial and polar non-axisymmetry effects from Eq.~\eqref{eq:delta-Psi}. However, the reason why this happens is that the proposed ppE waveform in~\cite{Loutrel:2022xok} for the non-axisymmetry scenario only considered secular corrections to the precession dynamics, i.e. secular effects within $(\epsilon,\alpha,\beta)$. A more thorough analysis of the precession dynamics in~\cite{Loutrel:2022ant} reveals that these quantities possess oscillatory corrections that couple to the non-axisymmetry parameters. In the course of our investigation herein, we discovered that these oscillatory effects are crucial for breaking the degeneracies among the non-axisymmetry parameters. 

To understand how to incorporate these oscillatory effects into the ppE waveform in Eq.~\eqref{eq:ppE-wave}, consider, for example, the oscillatory corrections to $\alpha$. In the limit $q_{\cm} \ll 1$, the full expression for the precession angle $\alpha$ takes the form
\begin{equation}
    \label{eq:alpha}
    \alpha = -\psi + \sum_{j=-2}^{2} C_{\alpha}^{(j)} e^{ij\psi} + {\cal{O}}(q_{\cm}^{2})\ ,
\end{equation}
where we provide the complex coefficients $C_{\alpha}^{(j)}$ in Appendix~\ref{app:coeffs}. It is important to note that, to leading order in $q_{\cm}$, the $j=\pm1$ terms in the above sum only depend on the axial parameters $(q_{1},a_{1})$, while the $j=\pm2$ terms only depend on the polar parameters $(q_{2},a_{2})$. The time domain waveform contains overtones in $\alpha$ of the form $e^{im'\alpha}$ which, when combined with Eq.~\eqref{eq:alpha}, can be written as
\begin{align}
    e^{im'\alpha} &= e^{im'\alpha_{0}} \times \sum_{j_{1}=-\infty}^{\infty} \left(-\frac{i C_{\alpha}^{(1)}}{|C_{\alpha}^{(1)}|} \right)^{j_{1}} J_{j_{1}}\left(2i|C_{\alpha}^{(1)}|\right) e^{ij_{1}\psi}
    \nn \\
    &\times \sum_{j_{2}=-\infty}^{\infty} \left(-\frac{i C_{\alpha}^{(2)}}{|C_{\alpha}^{(2)}|} \right)^{j_{2}} J_{j_{2}}\left(2i|C_{\alpha}^{(2)}|\right) e^{2ij_{2}\psi}\,,
    \nn \\
    &= e^{im'\alpha_{0}} \left[1 + C_{\alpha}^{(1)} e^{i\psi} + C_{\alpha}^{(2)} e^{2i\psi} + {\rm c.c.} + {\cal{O}}(q_{\cm}^{2})\right]
\end{align}
where $\alpha_{0} = -\psi + C_{\alpha}^{(0)}$, $J_{n}(x)$ is the Bessel function of the first kind, and c.c. is shorthand for complex conjugation of the preceding terms. The first equality above is provided for completeness and is obtained by using the generating functions of $J_{n}(x)$, while the second equality is obtained by linearizing in $q_{\cm}$. Thus, in the limit $q_{\cm} \ll 1$ which is most relevant to null tests of the no hair theorems, the axial non-axisymmetry effects only create nutation overtones with harmonic number $n=\pm1$, while the polar non-axisymmetry effects only create nutation overtones with $n=\pm2$. 

It is important to note that the analysis of~\cite{Loutrel:2022xok}, which only included secular effects in the precession dynamics, concluded that only the $n=0$ harmonics were relevant to the ppE waveforms in Eq.~\eqref{eq:ppE-wave}, due to the fact that when $q_{m} = 0$, there are no nutation harmonics arising from quadrupole-monopole couplings. The ppE waveform still takes the form of Eq.~\eqref{eq:ppE-wave} when including the nutation harmonics generated from oscillatory effects in the precession dynamics, but one must now consider harmonics with $n\neq0$. We, thus, propose a modification to the previous analysis of~\cite{Loutrel:2022xok}. For tests of \textit{axial} non-axisymmetric violations of the no hair theorems, one should consider $(m',n=0,\pm1)$ harmonics (three in total). On the other hand, for tests of \textit{polar} non-axisymmetry violations of the no hair theorems, one should consider $(m', n=0,\pm2)$ harmonics. One still has to identify which $m'$ harmonic is most relevant for null tests of the no hair theorems, and consequently, which will possess ppE parameters. However, this modification provides a clear distinction between the non-axisymmetry effects that the previous formalism lacks.

\subsection{Fisher Analysis}
\label{sec:fisher}

The goal of this study is to provide a projection of the upper bounds one can obtain on non-axisymmetric deformations of compact objects from GW observations with future detectors. A simplified method of doing this is provided by a Fisher analysis~\cite{Cutler:1994ys,Finn:1992wt,Poisson:1995ef}, which posits that the posterior on the waveform parameters $\theta^{a}$ is approximately a Gaussian distribution centered at the injected value of the parameters. The covariance matrix $\Sigma_{ab}$ is then approximated by the inverse of the Fisher matrix
\begin{equation}
    \label{eq:fisher-theta}
    \Gamma_{ab}^{\theta} = \left( \frac{\partial \tilde{h}}{\partial \theta^{a}} \Bigg| \frac{\partial \tilde{h}}{\partial \theta^{b}} \right)\,,
\end{equation}
where $\tilde{h}$ is the Fourier transform of the detector response, $(A|B)$ is the noise-weighted inner product
\begin{equation}
    \label{eq:inner-prod}
    (A | B) = 4 {\rm Re} \int_{f_{\rm low}}^{f_{\rm high}} \frac{df}{S_{n}(f)} A(f) B^{\dagger}(f)\,,
\end{equation}
with $\dagger$ corresponding to complex conjugation, and $S_{n}(f)$ the noise power density of the relevant detector. The limits of integration $[f_{\rm low}, f_{\rm high}]$ are determined by the sensitivity of the detector. For the analysis carried out here, we consider the constraints one can obtain from ET, CE, and LIGO, with the details of each detector given in Table~\ref{tab:detectors}. For $S_{n}(f)$ of each detector, we use the data provided in~\cite{Punturo-ET,ce-sens,Barsati-LIGO} (using ET-D for ET) and generate an interpolating function to sample at our desired frequency resolution. Since the waveform in Eq.~\eqref{eq:ppE-wave} only include the inspiral part of the coalescence, we cut the waveforms at $f_{\rm ISCO} = 1/(\pi M 6^{3/2})$, which fixes the upper limit of the integral in Eq.~\eqref{eq:inner-prod}. 

The uncertainty in the waveforms parameters $\Delta \theta^{a}$ is then given by the diagonal elements of the covariance matrix, specifically
\begin{equation}
    \Delta \theta^{a} = \sqrt{\Sigma_{aa}}\,,
\end{equation}
The inversion of the Fisher matrix is performed through a singular value decomposition (SVD), and we check the ratio of the minimum and maximum eigenvalues of $\Gamma_{ab}$ is larger than the limit of double precision accuracy. If this condition is violated, then the matrix inversion is ill-conditioned. With the Fisher matrix defined by Eq.~\eqref{eq:fisher-theta}, this is typically the case. To avoid this, we include the prior probability on each parameter in the following manner. Defining
\begin{align}
    \Gamma_{ab}^{0} = \text{diag}\left[(\delta \theta^{a})^{-2}\right]
\end{align}
with $\delta \theta^{a}$ the prior range on the parameter $\theta^{a}$, the covariance matrix is then
\begin{equation}
    \Sigma_{ab} = \left(\Gamma_{ab}^{\theta} + \Gamma_{ab}^{0}\right)^{-1}\,.
\end{equation}
Many of the waveform parameters are bounded, specifically $\eta \in [0,0.25]$, $(\theta_{N}, \theta_{S}, \beta_{0}) \in [0,\pi]$, $(\phi_{N}, \phi_{S}, \omega_{0}, \phi_{c}, \psi_{c}, \epsilon_{c}, a_{1,2}) \in [-\pi,\pi]$, $\chi_{Q} \in [-1,1]$, and $q_{1,2} \in [0,1]$. The priors on these parameters are then
\begin{align}
    \label{eq:priors}
    \delta \eta &= 0.25\,,
    \nn \\
    \delta \theta_{N} &= \delta \theta_{S} = 1\,,
    \nn \\
    \delta \beta_{0} &= \delta \phi_{N} = \delta \phi_{S} = \delta \omega_{0} = \delta \phi_{c} = \delta \psi_{c} = \delta \epsilon_{c} = \delta a_{\cm} = \pi\,,
    \nn \\
    \delta \chi_{Q} &= 1\,, \qquad \delta q_{\cm} = 1\,,
\end{align}
while we do not place priors on $(M, D_{L}, t_{c})$. 

The range of values for $q_{1,2}$ are determined by the approximations used to derive the waveform in Eq.~\eqref{eq:ppE-wave}, while the range on $\chi_{Q}$ comes from the following considerations. Up to a normalization factor, the mass quadrupole of a BH is $Q_{0} = - \chi^{2} m^{3}$, with $(m,\chi)$ the BH's mass and dimensionless spin. As a result, for GR BHs, %
\begin{equation}
    \label{eq:chiQ-to-spins}
    \chi_{Q} = -\eta_{2}^{2} \chi_{2}^{2} - \eta_{1}^{2} \chi_{1}^{2}\,.
\end{equation}
Since $\chi_{A} \in [0,1]$ and $\eta_{A} \in [0,1/2]$, the maximum value $\chi_{Q}$ can take is $-1$, which occurs in the limit of one of the masses become negligibly small. For equal mass BBHs, the maximum value is $-1/4$ due to limits on the mass ratio. In our analysis, we do not include spin contributions to the phase of the waveform. The above considerations from spinning BHs is merely to provide a reasonable injection value for $\chi_{Q}$ in our analysis in Sec.~\ref{sec:results1} \footnote{Note that, for this waveform model that only includes quadrupole-monopole precession, the individual spins are actually degenerate with each other, and only enter the waveform through $\chi_{Q}$ in Eq.~\eqref{eq:chiQ-to-spins}. Thus, it is not possible to get reasonable results on the uncertainty of the individual spins in a Fisher analysis using our waveform.}. For astrophysical/exotic compact objects, $\chi_{Q}$ is equation-of-state dependent, and can be larger than the values for BHs~\cite{Laarakkers:1997hb,Vaglio:2022flq}. However, we have found that the absolute errors on the waveform parameters depend only mildly on the injected value of $\chi_{Q}$, with their relative values changing by $\sim10$\%, provided it is restricted to the range of relevance of BBHs, i.e. $\chi_{Q} \in [-1,0]$.

\begin{table}
\centering
\begin{tabular} {c || c | c }
    Detector & $(f_{\rm low}, f_{\rm high})$ [Hz] & $S_{n}(f)$ \\
    \hline
    ET-D & (3, $f_{\rm ISCO}$) & ~\cite{Punturo-ET}  \\
    CE 40km & (3, $f_{\rm ISCO}$) & ~\cite{Evans:2023euw, ce-sens}  \\
    LIGO & (10,$f_{\rm ISCO}$) & ~\cite{Barsati-LIGO} 
\end{tabular}
\caption{Details of GW detectors used in this analysis.}
\label{tab:detectors}
\end{table}
%

\subsection{Averaging}
\label{sec:avg}

In~\cite{Loutrel:2022xok}, the results of the likelihood analysis were dependent on the various angles parametrizing the orientation of the binary. The Fisher analysis carried out in the following section also exhibit this behavior, in addition to a strong dependence on the sky location. A formal example of how this impacts the results of our Fisher analysis is provided in Sec.~\ref{sec:deg}. One could study plausible constraints on non-axisymmetry on a system-by-system basis, but when studying the impact of ppE deformations, it is better to study the constraints from an ensemble of systems. This could be done by repeating the Fisher analysis for a large number of systems, varying only the relevant parameters. However, this approach is inefficient since the number of required systems is typically large, and thus requires a non-trivial amount of computation time. To avoid this, we average over the relevant angle parameters, and describe here the averaging procedures used.

\subsubsection{Line of Sight Angles $\theta_{N},\phi_{N}$}

The waveforms described in Sec.~\ref{sec:waveform} depend on the angles parameterizing the line of sight in the binary's reference frame via
\begin{equation}
    \label{eq:spherical-harms}
    \tilde{h}(f) = \sum_{lm'} \tilde{h}_{lm'}(f) {_{-2}}Y_{lm'}(\theta_{N},\phi_{N})\,.
\end{equation}
As a result, the inner product in Eq.~\eqref{eq:inner-prod} averaged over the 2-sphere spanned by $(\theta_{N},\phi_{N})$ becomes
\begin{equation}
    \label{eq:avg-N}
    \langle A | B \rangle_{N} = \sum_{lm'} \left(A_{lm'} | B_{lm'}\right)\,,
\end{equation}
where $(A,B)$ are decomposed via Eq.~\eqref{eq:spherical-harms}, and we have used the orthogonality properties of the spin-weighted spherical harmonics. While the sum in Eq.~\eqref{eq:avg-N} only spans over $(l,m')$, there are additional summations within $(A_{lm'},B_{lm'})$ on the nutation index $n$ and the SUA index $k$. There is no orthogonality property for these harmonics, and as a result, the averaged inner product will mix different $(n,k)$ numbers. 

\subsubsection{Sky Location Angles $\theta_{S},\phi_{S}$}

GW detectors are well known to not have equal sensitivity across the celestial sphere. As we are primarily concerned with reasonable order of magnitude estimates on the uncertainties of the non-axisymmetry parameters, we sky average the inner product in Eq.~\eqref{eq:inner-prod}. For precessing signals, the sky averaging is complicated by the fact that the polarization angle in Eq.~\eqref{eq:pol-angle} is parametrized by the sky location through $\vec{N}$ or $\hat{z}$, depending on whether one computes this quantity in the rest frame of the binary or the detector. However, the TT gauge that specifies the observable waveform polarizations $h_{+,\times}$ is only known up to an arbitrary rotation of the polarization basis, corresponding to the residual gauge freedom associated with the transverse subspace~\cite{PoissonWill:sec:11.2}. As a result, $\psi_{\rm pol}$ in Eq.~\eqref{eq:pol-angle} can be generalized by introducing a constant shift $\psi_{0}$. We perform the sky averaging with respect to $\psi_{0}$, and thus
\begin{equation}
    \langle F_{+}^{2} \rangle_{S} = 1/5 = \langle F_{\times}^{2} \rangle_{S}\,, \qquad \langle F_{+} F_{\times} \rangle_{S} = 0\,,
\end{equation}
where
\begin{equation}
    \langle F \rangle_{S} = \frac{1}{8\pi^{2}} \int_{0}^{\pi} d\theta_{S} \sin\theta_{S} \int_{0}^{2\pi} d\phi_{S} \int_{0}^{2\pi} d\psi_{0} F(\theta_{S}, \phi_{S}, \psi_{0})\,,
\end{equation}
for an arbitrary function of the sky location $F$. Combining this procedure with the average over the line of sight, the Fisher matrix is now defined in terms of the double averaged inner product $\langle A | B \rangle_{NS}$.

\section{Constraints on Violations of the Black Hole No Hair Theorems}
\label{sec:results1}

In this section, we consider the plausibility of constraining the no hair theorems with GWs emitted during the inspiral of BBH coalescences. Since BHs are spheroidal by the no hair theorems, $q_{\cm} = 0$ for BHs in GR~\footnote{We only consider BHs as they are described by GR, and with no electric charge, i.e. Kerr BHs.}. More specifically, it is the components of the mass quadrupole $Q_{\cm\neq0}$ that vanish. As a result, the non-axisymmetry argument parameters $a_{\cm}$ are technically not well-defined in the BH limit, due to the mapping in Eq.~\eqref{eq:non-axi-defs}. For BBH events in ET and CE, we normalize the luminosity distance such that the SNR $\rho = \sqrt{(h|h)}$ is equal to one thousand, while for LIGO events we normalize such that $\rho=100$. Note that the loudest BBH events in ET are expected to have SNRs of order $\mathcal{O}(10^3)$, see for example Fig. 5 in~\cite{Branchesi:2023mws}. Therefore the normalization choice SNR=1000 serves as a good indicator for the upper bounds one can obtain with ``golden events.''

\subsection{Degeneracies in Non-Axisymmetry Parameters $q_{1,2}$}
\label{sec:deg}

As a formal display of the impact of the degeneracies among the non-axisymmetry parameters in Eq.~\eqref{eq:delta-Psi}, we perform a Fisher analysis on a $(10,10) M_{\odot}$ binary. For this example, we do not average over the sky location and direction of propagation, and we do not include the nutation effects proposed in Sec.~\ref{sec:avg}. This is merely to provide an example of the pitfalls of the ppE analysis that only use secular effects in the precession dynamics, as was proposed in~\cite{Loutrel:2022xok}.

\begin{figure*}[hbt!]
    \centering
    \includegraphics[width=\columnwidth]{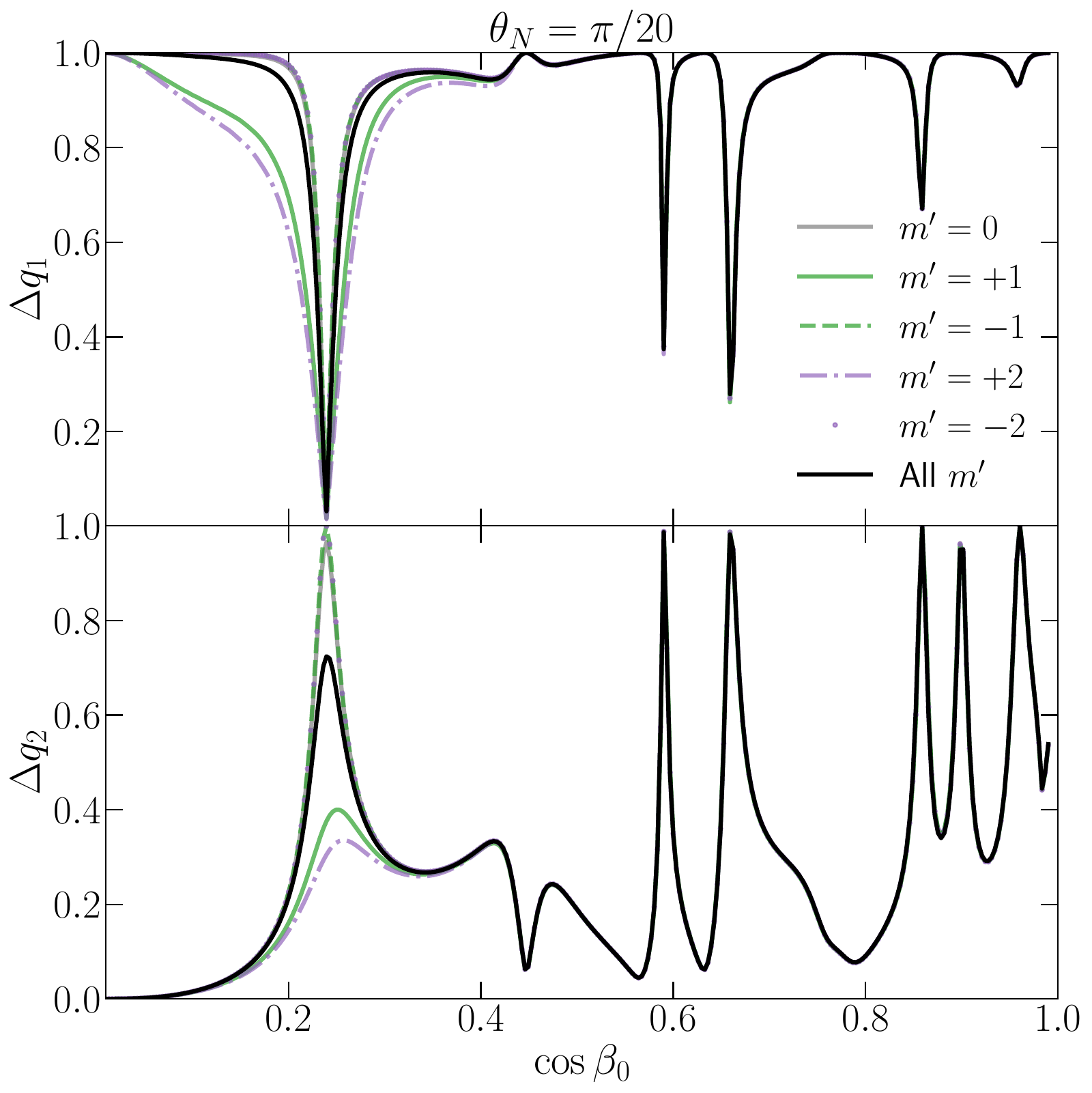}
    \includegraphics[width=\columnwidth]{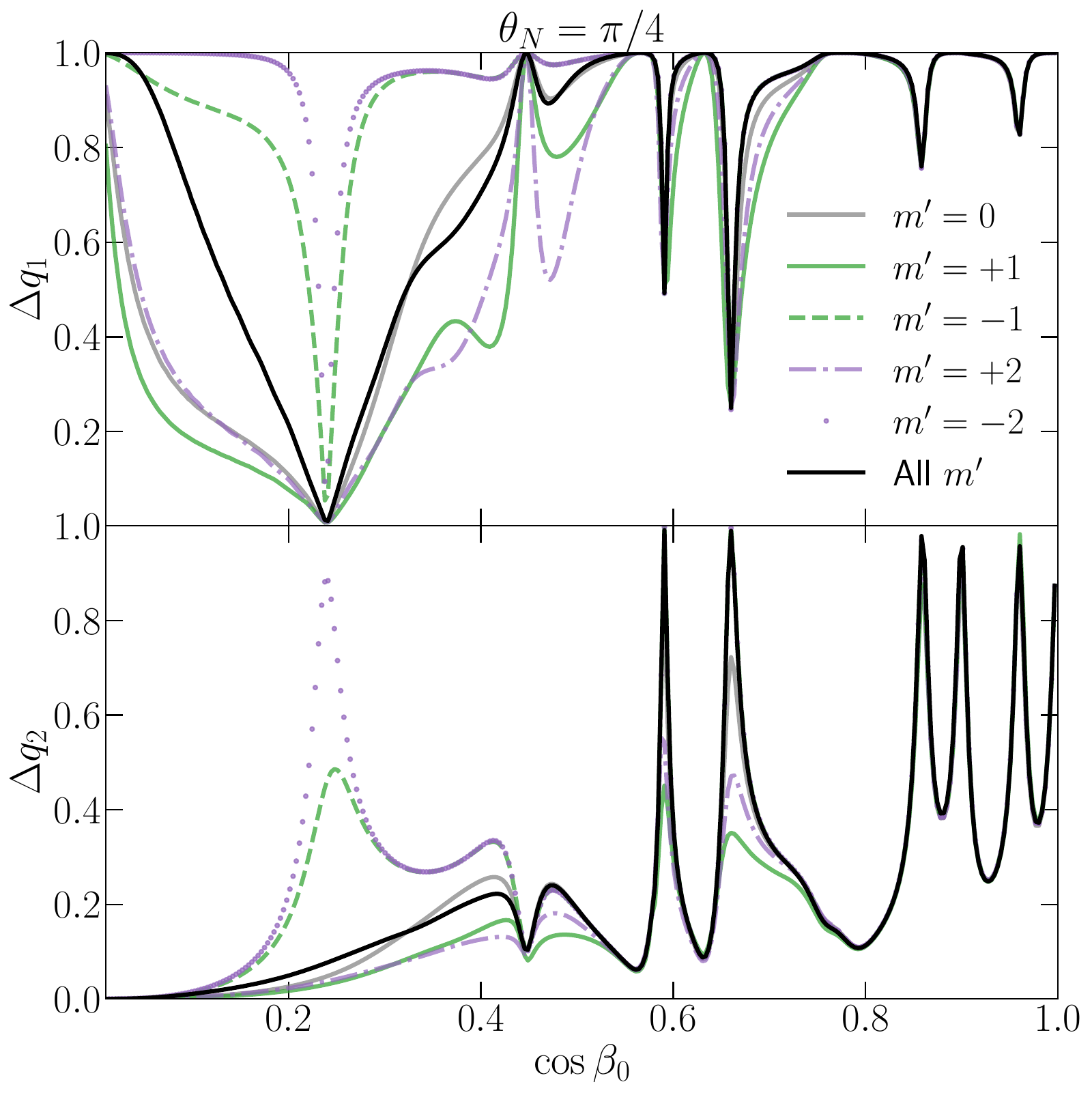}
    \caption{Uncertainty on the non-axisymmetry modulus parameters $q_{1}$ (top panel) and $q_{2}$ (bottom panel) as a function of $\cos\beta_{0}$, for line of sight angles $\theta_{N} = \pi/20$ (left) and $\pi/4$ (right). The uncertainty is computed from the ppE waveform in Eq.~\eqref{eq:ppE-wave} and varying the $(m',n=0)$ harmonic which have ppE deformations, and neglecting the nutation overtones. The black lines provide the uncertainty with all $(m', n=0)$ harmonics having ppE deformations, as was proposed in~\cite{Loutrel:2022xok}, while the colored lines limit the ppE deformations to only the specific $m'$-harmonic ($n=0$). Which harmonic produces the lowest $\Delta q_{\cm}$, and thus strongest constraints on non-axisymmetry, will generally depend on the line of sight angles $(\theta_{N},\phi_{N})$, the sky location $(\theta_{S}, \phi_{S})$, and orientation angles $(\beta_{0}, \omega_{0})$.}
    \label{fig:ppE}
\end{figure*}

Fig.~\ref{fig:ppE} provides the uncertainty in the axial $q_{1}$ (top) and polar $q_{2}$ (bottom) parameters for two values of the direction of propagation $(\theta_{N} = \pi/20, \phi_{N} = 0)$ (left) and $(\theta_{N} = \pi/4, \phi_{N} = 0)$ (right). The uncertainties are plotted as a function of the initial inclination angle, which is achieved by sampling $\beta_{0}$ over three hundred values, and repeating the Fisher analysis for each value. Each line corresponds to which $m'$ harmonic contains the ppE parameters, which are given in Appendix~\ref{app:coeffs}. The solid black line is the case when all five $m'$ harmonics possess ppE parameters. 

There are a few things to note here. First, the degeneracy between $q_{1}$ and $q_{2}$ arising from Eq.~\eqref{eq:delta-Psi} is readily apparent. In Fig.~\ref{fig:ppE}, when $\Delta q_{1}$ is large (near unity), $\Delta q_{2}$ is small, and vice versa. This is due to the fact that, practically, this analysis only places a bound upon the 2PN deviation in Eq.~\eqref{eq:delta-Psi}, and the uncertainties in the underlying parameters, namely $q_{1,2}$, are then defined by an error ellipse generated from this bound. The one exception to this occurs near $\cos\beta_{0} \sim 0.22$, which is a result of the correlations reaching a minimum with respect to $\beta_{0}$. However, the impact of this maximum is also dependent on which harmonic(s) possess ppE parameters.

Second, which harmonic gives the best constraint on the non-axisymmetry parameters $q_{1,2}$ varies depending on the values of $\theta_{N}$ and $\beta_{0}$. Repeating this analysis while varying $(\phi_{N}, \theta_{S}, \phi_{S})$ reveals similar dependence on these quantities. This is not unexpected since a similar behavior was found in the simplified likelihood analysis of~\cite{Loutrel:2022xok}. However, that analysis found a clear dependence on $\beta_{0}$ as can be seen in Fig.~3 therein, which we do not find here. The reason for this appears to be that the dependence found therein was dominated by the amplitudes' dependence on $\beta_{0}$, where here the uncertainties are primarily dominated by the phase, which has a complicated dependence on $\beta_{0}$, creating strong covariances with the non-axisymmetry parameters. 

Lastly, \textit{including ppE parameters in multiple $m'$ harmonics does not necessarily improve bounds on the non-axisymmetry parameters}. Thus, one needs to identify which $m'$ harmonic places the strongest bounds. Due to the dependence on sky location and direction of propagation, we perform the averaging described in Sec.~\ref{sec:avg} for the remainder of this work. Naively, this seems to also contradict the proposal in Sec.~\ref{sec:osc}. However, this conclusion is drawn from the secular analysis of~\cite{Loutrel:2022xok} and thus only applies to the $m'$ harmonics. One still needs to include the nutation harmonics in order to disentagle the axial and polar non-axisymmetry effects. 

\subsection{Analysis of ppE Corrections}
\label{sec:ppE}

We now use the results of our Fisher analysis to identify which $m'$ harmonic is most relevant for constraining, separately, the axial and polar non-axisymmetry effects. For the axial case, we vary which $m'$ harmonic has the proposed ppE parameters in Appendix~\ref{app:coeffs} and only allow the chosen $m'$ harmonic to have the $n=\pm1$ overtones with the appropriate ppE parameters. The polar case follows the same procedure, but with only the $n=\pm2$ overtones. Rather than selecting the other parameters of BBH systems at random, we choose parameters corresponding to a few of the events in the GWTC-3 catalog~\cite{GWTC3}. We do not perform our analysis for all of the GWTC-3 events, but choose a subset of events whose marginalized posteriors on $\chi_{p}$ peak away from zero (see Fig. 7 of~\cite{GWTC3}). It is important to note that these events plausibly correspond to the coalescence of spinning BBH systems, and thus, any precession within the observed LIGO signals would be due to spin precession, not the quadrupole precession being considered here. We simply pick these systems as useful benchmarks for our analysis. The parameters of the selected events are displayed in Table~\ref{tab_et}. For all of the cases, we choose $\chi_{Q}=0.25$. Note that one of these systems, namely GW200115, is plausibly a NSBH binary. However, we include it in our analysis to cover systems with a broader range of total masses, and consider it as another BBH signal.

\begin{figure*}[hbt!]
    \includegraphics[width=\columnwidth]{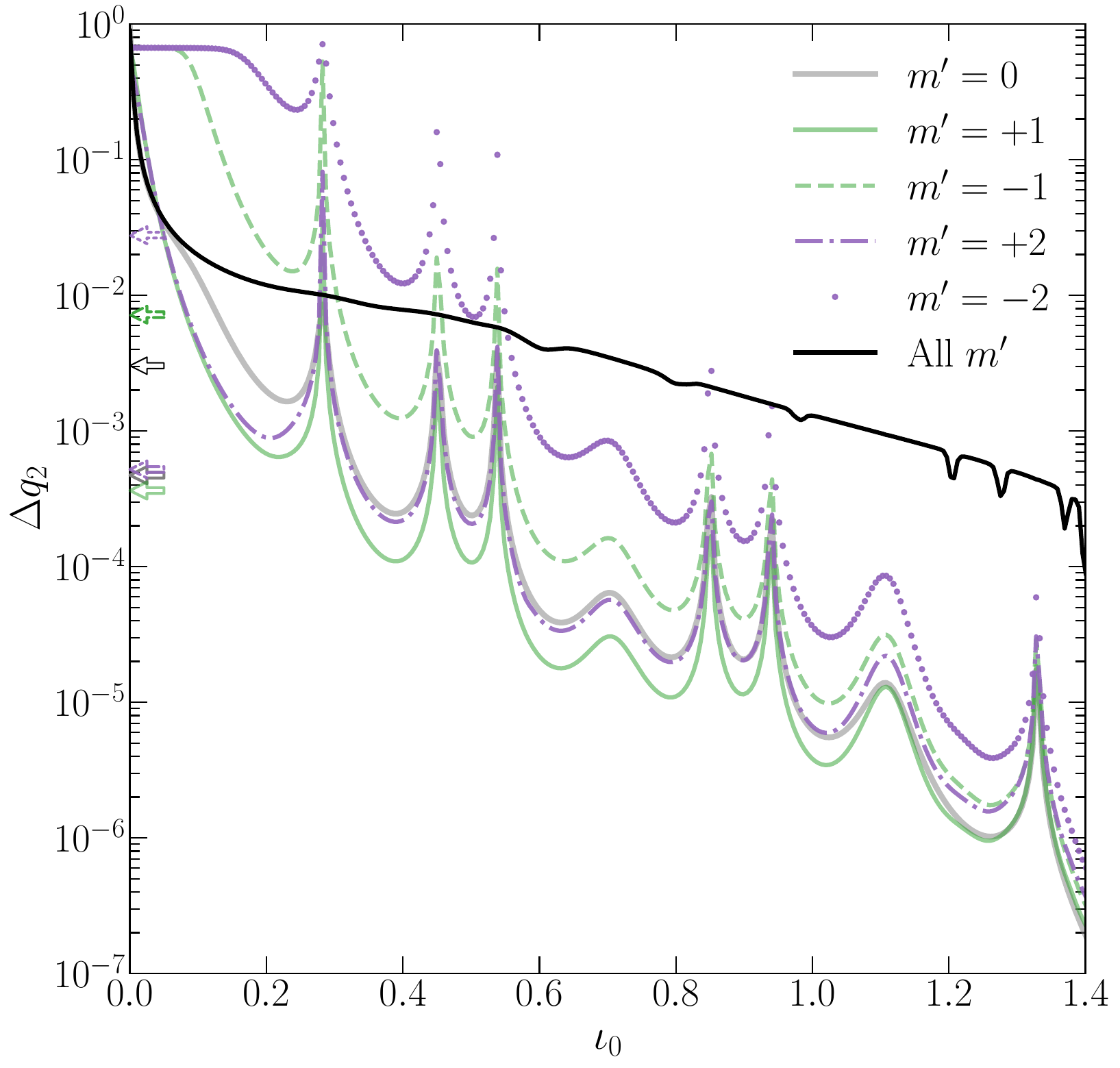}
    \includegraphics[width=\columnwidth]{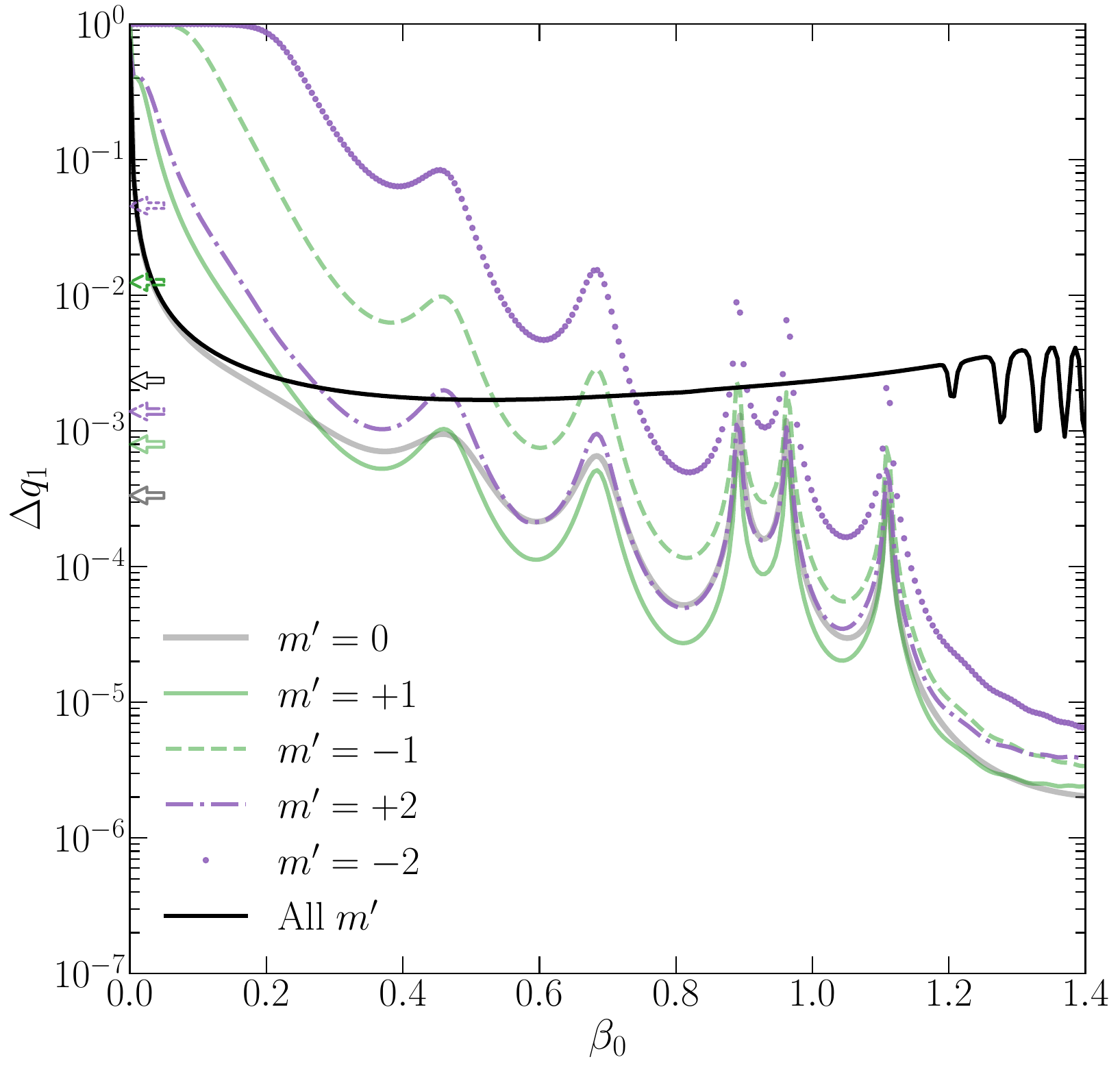}
    \caption{Left: Uncertainty for a GW200316-like binary on the polar argument parameter $q_{2}$ as a function of the initial inclination angle $\beta_{0}$. In this case the ppE waveforms have been modified to include the $n=0,\pm2$ harmonics, corresponding to nutation overtones due to polar asymmetries. Each line corresponds to scenarios with different $m'$ harmonics containing ppE parameters, specifically $m'=0$ (gray), $m'=\pm1$ (green solid/dashed), and $m'=\pm2$ (purple dot-dashed/dotted). The black solid line corresponds to the scenario where all five $m'$ harmonics contain the non-axisymmetry deformations. The arrows on the y-axis show the value of the uncertainty for each scenario averaged over $\beta_{0}$. Right: The same analysis, but for the axial argument parameter $q_{1}$. The ppE waveforms are modified to contain the $n=0,\pm1$ nutation overtones, corresponding to axial asymmetries.}
    \label{fig:gw200316}
\end{figure*}

Similar to the analysis of Sec.~\ref{sec:deg}, we compute the uncertainties for each event as a function of $\beta_{0}$. For all of the events, the uncertainties on the argument parameters $a_{1,2}$ are dominated by the priors, i.e. $\Delta a_{1,2} \sim \pi$. This is not a problem from the standpoint of null tests of the no hair theorems. The argument parameters $a_{\cm}$ determine the orientation of the body's non-axisymmetric deformations, while the modulus parameters $q_{\cm}$ describe how large the deformations are. The uncertainties on the modulus parameters $q_{1,2}$ are strongly dependent on $\beta_{0}$. An example of this is shown in Fig.~\ref{fig:gw200316} for the system GW200316. Since we are now including the nutation overtones and are considering the constraints on $q_{1,2}$ separately, we plot the uncertainties separately.

Fig.~\ref{fig:gw200316} shows that the degeneracies between $q_{1,2}$ and $\beta_{0}$ are still present, but the inclusion of the nutation overtones has allowed greater qualitative delineation in the uncertainties due to each harmonic. Note that, unlike the case in Fig.~\ref{fig:ppE}, including ppE corrections in all precesion overtones eliminates the sharp features arising from the covariance between the $q_{m}$'s and $\beta_{0}$. This results from the Fisher matrix now containing cross terms between harmonics of different $n$ number when computing the inner product, which smear out these features. Further, because we have averaged over sky location and line of sight orientation, it is much easier to identify which harmonic is most relevant for a ppE-style analysis. For completeness, we average each case over $\beta_{0}$, specifically
\begin{equation}
    \label{eq:beta-avg}
    \langle \Delta \theta^{a} \rangle_{\beta_{0}} = \frac{\int_{0}^{\pi/2} d\beta_{0} \sin \beta_{0} \Delta \theta^{a}}{\int_{0}^{\pi/2} d\beta_{0} \sin\beta_{0}}\,.
\end{equation}
The averages are indicated by the arrows on the y-axis in Fig.~\ref{fig:gw200316}. For a GW200316-like binary, the $m'=0$ harmonic provides the strongest constraints for axial asymmetries, while the $m'=+1$ harmonic is dominant for polar asymmetries. We repeat this analysis for the remaining systems selected from the GWTC-3 catalog in Table~\ref{tab_et}, and find that this result is consistent.

Before continuing, we note a practical point about the upper limit of integration in Eq.~\eqref{eq:beta-avg}. When performing the Fisher analysis for $q_{2}$, the matrix inversion generally becomes ill-conditioned for $\beta_{0} \gtrsim 1.4$ due to the waveform derivatives diverging as $\beta_{0} \rightarrow \pi/2$. This is due to a known breakdown of the solutions of the precession dynamics, which was pointed out in Eq.~(35) of~\cite{Loutrel:2022xok}. Thus, when performing the average of our results for $q_{2}$, we replace the upper limit of integration with $\beta_{0} = 1.4$. This same issue does not happen in our Fisher analysis for $q_{1}$, due to the divergences not being present in this case. As a result, we have tested how much replacing the upper limit of integration in Eq.~\eqref{eq:beta-avg} with $\beta_{0} = 1.4$ in our analysis of ppE corrections and upper bounds for $q_{1}$. We find that the uncertainties are larger when replacing the upper limit of integration by about $\sim10\%$, indicating that using $\beta_{0} = 1.4$ does not impact our analysis of the ppE harmonic, and that the upper bounds (which are discussed in Sec.~\ref{sec:bounds}) listed in Table~\ref{tab_et}-\ref{tab_ligo} are conservative values.

To further narrow down the proper structure of the waveform's phase, we test whether the SUA function $u_{k}^{\rm GR}(f)$ in Eq.~\eqref{eq:delta-Psi} is needed, or can be replaced with its leading PN order expansion, i.e. $u_{k}^{\rm GR}(f) \sim \tilde{u} + {\cal{O}}(\tilde{u}^{7/2})$. Note that, because the SUA summation index $k$ enters in the ${\cal{O}}(\tilde{u}^{7/2})$ terms, this test can be achieved by setting $k_{\rm max} = 0$. The upper bounds that we discuss in Sec.~\ref{sec:bounds} and report in Table~\ref{tab_et} are all achieved with $k_{\rm max}=1$. When comparing to the scenario with SUA correction turned off, we find that the upper bounds on the asymmetry parameters $q_{\cm}$ only change negligibly, specifically beyond the third decimal point for the results in Table~\ref{tab_et}. As a result, one can replace $u_{k}^{\rm GR}(f) \rightarrow \tilde{u}$ in Eq.~\eqref{eq:delta-Psi} without loss of accuracy, and the ppE phase corrections will now take the standard form of the original ppE formalism. This shouldn't be surprising since it has already been found that constraints on modified theories of gravity with the ppE formalism are robust to higher PN order effects within GR~\cite{Perkins:2022fhr}.

From this analysis, we may identify the ``correct" ppE parameters for non-axisymmetric mass quadrupole effects. For axial asymmetries, the correct ppE amplitude parameters are 
\begin{align}
    \label{eq:bppE-ax}
    \cb_{0,n}^{\rm axial} &= \frac{\sqrt{5\pi}}{4} q_{1} \chi_{Q} \left[ {\cal{U}}_{10} - \frac{3}{4} \cos\beta_{0} \Delta\Omega_{\epsilon}^{(1)} 
    \right.
    \nn \\
    &\left.
    + \frac{3}{8} \left(n -2\Omega_{\epsilon}^{(0)}\right)\sin\Delta_{a} \tan\beta_{0}\right]\,,
    \\
    \ca_{0,0}^{\rm axial} &= 2 q_{1} \cot\beta_{0} \sin\Delta_{a}\,,
    \\
    \ca_{0,\pm1}^{\rm axial} &= i q_{1} \left(2\sec\beta_{0} \mp e^{\mp i\Delta_{a}} \cos\beta_{0} \cot\beta_{0} \right)\,,
\end{align}
with $\Delta_{a} = a_{1} - a_{2}$, while the correct ppE amplitude parameters for polar asymmetries are 
\begin{align}
    \cb_{+1,n}^{\rm polar} &= \frac{\sqrt{5\pi}}{4} q_{2} \chi_{Q} \left[{\cal{U}}_{01} - \frac{3}{4} \cos\beta_{0} \Delta\Omega_{\epsilon}^{(2)}
    \right.
    \nn \\
    &\left.
    + \frac{3}{8} \left(n - 1 - 2\Omega_{\epsilon}^{(0)}\right)\tan^{2}\beta_{0}\right]\,,
    \\
    \label{eq:appE-pol}
    \ca_{+1,0}^{\rm polar} &= - q_{2} \tan\beta_{0} \,,
    \\
    \ca_{+1,-2}^{\rm polar} &= \frac{q_{2}}{8} \left(5 - \sec\beta_{0}\right)\left(1 - \sec\beta_{0}\right)
    \\
    \ca_{+1,+2}^{\rm polar} &= \frac{q_{2}}{8} \left(3 - \sec\beta_{0}\right)\left(1 + \sec\beta_{0}\right).
\end{align}
Note that the factors of $[\Omega_{\epsilon}^{(0)},\Delta\Omega_{\epsilon}^{(1,2)}]$ come from correcting a typo in Eq.~(120) of~\cite{Loutrel:2022xok}. These quantities are explicitly given in Eqs.~\eqref{eq:dOm1}-\eqref{eq:dOm2}. For both scenarios, the ppE exponent parameters are the same, namely $b_{m',n}^{\rm ppE} = -1$ and $a_{m',n}^{\rm ppE} = 0$. Note that the axial ppE parameters only hold for $n=0,\pm1$, while the polar ppE parameters only hold for $n=0,\pm2$.

Finally, we point out that, unlike the original ppE formalism for non-precessing quasi-circular binaries, ppE parameters \textit{are} needed in multiple harmonics. We reiterate that the reason for this is to break the degeneracy between axial and polar axisymmetry that appears at 2PN order in the waveform's phase. This is not the case for spin-precessing binaries in Chern-Simons gravity~\cite{Loutrel:2022xok}, so it is plausible that the need for ppE parameters in multiple waveform harmonics is unique to the physical scenario investigated here, or at least unique to the subset of physical scenarios where multiple non-GR effects appear at the same PN order in the phase. However, we can only speculate on this point at this stage. This issue needs to be evaluated on a case-by-case basis.

\subsection{Constraints on Non-Axisymmetry}
\label{sec:bounds}

Having identified the relevant ppE parameters for quadrupolar non-axisymmetry effects, we now report the plausible constraints on these effects with future generation detectors. Table~\ref{tab_et} provides the constraints from the six systems selected from the GWTC-3 catalog, but using the ET detector. Recall that all signals are normalized such that $\rho=1000$. These results suggest that, for some of the loudest BBH events in ET, constraints on non-axisymmetric violations of the no hair theorems should be possible to $\sim 10^{-3}-10^{-4}$, subject to the assumptions of our analysis, which we discuss in Sec.~\ref{sec:disc}. For completeness, we also include the constraints one would obtain from a non-precessing system with the same masses in the fifth and seventh columns of Table~\ref{tab_et}. This corresponds to the case $\beta_{0} = 0$, and thus, we do not average over the initial inclination angle. The constraints from the non-precessing systems are largely dominated by the priors in Eq.~\eqref{eq:priors}, and thus, \textit{one can only place weak constraints on non-axisymmetry with non-precessing binaries. This highlights the importance of precession dynamics in placing stringent constraints on violation of the no hair theorems}.

We also investigate the projected upper bounds for the same systems with the CE and LIGO detectors in Tables~\ref{tab_ce}-\ref{tab_ligo}, respectively. The sources in the LIGO detector are normalized to have SNR $\rho=100$, while for CE $\rho=1000$. The upper bounds for precessing binaries in the CE detector are up to $\sim 50\%$ better than the ET upper bounds, while the LIGO upper bounds are $\sim2$ orders of magnitude weaker. Further, much like the ET sources, one can only place weak constraints on non-axisymmetry with non-precessing binaries with CE sources. For LIGO sources, the lower SNR prevents any meaningful constraints in the non-precessing case.

The greater than two order of magnitude improvement in the uncertainties for CE and ET compared to LIGO is rather surprising, considering that the SNR only increases by an order of magnitude. The uncertainties in a Fisher analysis scale inversely with the SNR, so naively, one might expect only about an order of magnitude improvement. However, CE and ET have significantly better sensitivity at lower frequencies than LIGO. When we adjust $f_{\rm low} = 50\,{\rm Hz}$ and normalize the SNR in all detectors to be one hundred, the uncertainties we obtain for all three detectors are identical to within the numerical error in our computation. Thus, the greater than one order of magnitude improvement for CE and ET signals compared to LIGO is a result of the improved sensitivity at low frequencies for third generation~(3G) detectors.

We have also investigated the improvement in the uncertainties on remaining (non-quadrupole) waveform parameters. For sources with the same SNR, the greater sensitivity at lower frequencies for 3G detectors improves the uncertainties on all of the parameters by $\sim {\rm few}\times 10\%$, with the exceptions of the luminosity distance $D_{L}$ and inclination angle $\beta_{0}$, which improve by an order of magnitude compared to LIGO. Further, unlike standard non-precessing TaylorF2 waveforms~\cite{Buonanno:2009zt} where the luminosity distance and inclination angle are completely degenerate with one another, the correlation coefficient between these parameters is $~10^{-2}$ for $\beta_{0} < \pi/3$, indicating the importance of precession for breaking degeneracies even for parameters within the GR sector of waveform models.
We conclude that the superior low-frequency sensitivity of 3G detectors is important to measure precisely some of the leading waveform parameters. In turn, this helps improve the degeneracy-breaking with high-PN corrections such as anomalous multipole moments, which are significantly better constrained by 3G detectors with respect to LIGO.

\begin{table*}[hbt!]
\centering
\begin{tabular} {c || c | c | c | c | c | c}
    Event & $M[M_{\odot}]$ & $\eta$ & $\Delta q_{1}$ (Prec) & $\Delta q_{1}$ (No-prec) & $\Delta q_{2}$ (Prec) & $\Delta q_{2}$ (No-prec) \\
    \hline
    GW191113 & 34.5 & 0.143 & $3.51\times10^{-4}$ & $1.00$ & $4.51\times10^{-4}$ & $0.990$\\
    GW191129 & 17.5 & 0.234 & $3.28\times10^{-4}$ & $1.00$ & $3.39\times10^{-4}$& $0.566$\\
    GW191216 & 19.8 & 0.238 & $3.33\times10^{-4}$ & $1.00$ & $3.54\times10^{-4}$ & $0.609$\\
    GW200115 & 7.40 & 0.155 & $3.02\times10^{-4}$ & $1.00$ & $2.69\times10^{-4}$ & $0.797$\\
    GW200210 & 24.1 & 0.0936 & $3.39\times10^{-4}$ & $1.00$ & $4.21\times10^{-4}$ & $0.978$\\
    GW200316 & 21.2 & 0.229 & $3.35\times10^{-4}$ & $1.00$ & $3.65\times10^{-4}$ & $0.669$
\end{tabular}
\caption{Projected ET upper bounds on the non-axisymmetry modulus parameters $q_{1,2}$ for selected GWTC-3 events. All sources are normalized such that $\rho = 1000$. The fourth and sixth columns provide the upper bounds for precessing systems, which are averaged over the initial inclination angle $\beta_{0}$, while the fifth and seventh columns provide the upper bounds for non-precessing systems $(\beta_{0} = 0)$.}
\label{tab_et}
\end{table*}
\begin{table*}[hbt!]
\centering
\begin{tabular} {c || c | c | c | c | c | c}
    Event & $M[M_{\odot}]$ & $\eta$ & $\Delta q_{1}$ (Prec) & $\Delta q_{1}$ (No-prec) & $\Delta q_{2}$ (Prec) & $\Delta q_{2}$ (No-prec) \\
    \hline
    GW191113 & 34.5 & 0.143 & $1.57\times10^{-4}$ & $1.00$ & $3.10\times10^{-4}$ & $0.992$\\
    GW191129 & 17.5 & 0.234 & $1.51\times10^{-4}$ & $1.00$ & $2.24\times10^{-4}$ & $0.600$\\
    GW191216 & 19.8 & 0.238 & $1.53\times10^{-4}$ & $1.00$ & $2.34\times10^{-4}$ & $0.655$\\
    GW200115 & 7.40 & 0.155 & $1.38\times10^{-4}$ & $1.00$ & $1.86\times10^{-4}$ & $0.830$ \\
    GW200210 & 24.1 & 0.0936 & $1.55\times10^{-4}$ & $1.00$ & $3.17\times10^{-4}$ & $0.986$\\
    GW200316 & 21.2 & 0.229 & $1.53\times10^{-4}$ & $1.00$ & $2.41\times10^{-4}$ & $0.719$
\end{tabular}
\caption{Projected CE upper bounds on the non-axisymmetry modulus parameters $q_{1,2}$ for selected GWTC-3 events. All sources are normalized such that $\rho = 1000$. The fourth and sixth columns provide the upper bounds for precessing system, which are averaged over the initial inclination angle $\beta_{0}$, while the fifth and seventh columns provide the upper bounds for non-precessing systems $(\beta_{0} = 0)$.}
\label{tab_ce}
\end{table*}
\begin{table*}[hbt!]
\centering
\begin{tabular} {c || c | c | c | c | c | c}
    Event & $M[M_{\odot}]$ & $\eta$ & $\Delta q_{1}$ (Prec) & $\Delta q_{1}$ (No-prec) & $\Delta q_{2}$ (Prec) & $\Delta q_{2}$ (No-prec) \\
    \hline
    GW191113 & 34.5 & 0.143 & $5.26\times10^{-2}$ & $1.00$ & $3.05\times10^{-2}$ & $1.00$\\
    GW191129 & 17.5 & 0.234 & $4.70\times10^{-2}$ & $1.00$ & $2.52\times10^{-2}$ & $1.00$\\
    GW191216 & 19.8 & 0.238 & $4.78\times10^{-2}$ & $1.00$ & $2.60\times10^{-2}$ & $1.00$ \\
    GW200115 & 7.40 & 0.155 & $4.26\times10^{-2}$ & $1.00$ & $2.05\times10^{-2}$ & $1.00$\\
    GW200210 & 24.1 & 0.0936 & $4.93\times10^{-2}$ & $1.00$ & $2.86\times10^{-2}$ & $1.00$\\
    GW200316 & 21.2 & 0.229 & $4.83\times10^{-2}$ & $1.00$ & $2.64\times10^{-2}$ & $1.00$
\end{tabular}
\caption{Projected LIGO upper bounds on the non-axisymmetry modulus parameters $q_{1,2}$ for selected GWTC-3 events. All sources are normalized such that $\rho = 100$. The fourth and sixth columns provide the upper bounds for precessing system, which are averaged over the initial inclination angle $\beta_{0}$, while the fifth and seventh columns provide the upper bounds for non-precessing systems $(\beta_{0} = 0)$.}
\label{tab_ligo}
\end{table*}

There are a few important things to note from these results. First, the upper bounds in Table~\ref{tab_et} are approximately three orders of magnitude smaller than those obtained from the analysis in Fig.~\ref{fig:ppE}. The primary reason for this is that analysis of Fig.~\ref{fig:ppE} only used the secular components of the precession quantities $[\alpha,\beta,\epsilon]$, which produces a strong degeneracy between the asymmetry parameters $q_{\cm}$. Introducing the oscillatory corrections to these generates nutation overtones in the waveform, which break the degeneracy in such a way that one can consider placing upper bounds on $q_{1}$ and $q_{2}$ separately within the context of a null hypothesis test.

Second, the bounds on $q_{1,2}$ are not bounds on the structure of the individual BHs in the binary. This is a result of the fact that the dynamics of the binary at leading PN order in the quadrupole-monopole interaction are only dependent on the effective mass quadrupole defined in Eq.~\eqref{eq:Qeff}. For the nearly equal mass binaries in Table~\ref{tab_et}, one will need to include higher PN order corrections to the quadrupole-monopole interaction to break this degeneracy. Alternatively, one could study this same scenario in the EMRI limit, where the quadrupole moment of the small body becomes negligibly small compared to the large body. We leave this to future work.

\section{Discussion}
\label{sec:disc}

In this work, we have considered, for the first time, the possibility of directly probing violations of the no hair theorems due to axisymmetry breaking using GW observations. Constraints on these violations with BBH events in current and future ground-based detectors are promising for precessing binaries, but are lacking for non-precessing events. This highlights the importance of tests of GR with GWs from precessing binaries.

It is important, however, to point out a pitfall of this analysis, and of the previous study in~\cite{Loutrel:2022ant}. In our analysis, we have neglected the spins of the BHs, and assumed that the precession is completely controlled by the quadrupole-monopole interaction. For realistic signals, this is not true, and relativistic spin precession is likely the dominant effect. While there has been much work in understanding spin precession in the context of BHs and other spheroidal compact objects, it remains to be understood how spin precession is modified when compact objects have broken axisymmetry. 

Further, the inclusion of spin precession effects in the waveforms will necessarily introduce additional parameters to the list given in Sec.~\ref{sec:waveform}. More specifically, the three components of the individual spin vectors must be included, introducing six additional parameters to recover. Generally, introducing additional parameters into the waveform models will weaken constraints on non-GR parameters, but precession can aid in recovering stringent constraints (see for example~\cite{Stavridis:2009mb}). A thorough investigation of this goes outside of the scope of this work, but is greatly needed.

Another avenue for future directions is to consider possible constraints with observations from space-based detectors. As is discussed in Sec.~\ref{sec:bounds}, EMRIs provide a means of breaking the degeneracy in Eq.~\eqref{eq:Qeff}, and would allow direct constraints on the multipolar structure of supermassive BHs. In addition, massive BBH systems detected by ground-based detectors, such as GW150914~\cite{gw150914} and GW190521~\cite{gw190521}, may also be detected years before merger by space-based observatories like LISA~\cite{Amaro-Seoane:2012vvq}. One could thus consider the possibility of performing multiband tests of the no hair theorems with such systems.


\acknowledgements

N.L. \& P.P. acknowledge financial support provided under the European Union's H2020 ERC, Starting Grant agreement no.~DarkGRA--757480. N.L. \& P.P. also acknowledge support under the MIUR PRIN and FARE programmes (GW-NEXT, CUP: B84I20000100001), and from the Amaldi Research Center funded by the MIUR program ``Dipartimento di Eccellenza'' (CUP: B81I18001170001). 
N.L. is also supported by ERC Starting Grant No.~945155--GWmining, 
Cariplo Foundation Grant No.~2021-0555, MUR PRIN Grant No.~2022-Z9X4XS, 
and the ICSC National Research Centre funded by NextGenerationEU. 
R.B. acknowledges financial support provided by FCT/Portugal under the Scientific Employment Stimulus -- Individual Call -- 2020.00470.CEECIND, and under project No. 2022.01324.PTDC.

\appendix

\section{Coefficients of ppE and Oscillatory Effects}
\label{app:coeffs}

The precessing ppE waveforms are given in Eq.~\eqref{eq:ppE-wave} with amplitude coefficients $(\ca_{\slashed{K}}^{\rm ppE}, \cb_{\slashed{K}}^{\rm ppE})$ and exponent coefficients $(a_{\slashed{K}}^{\rm ppE},b_{\slashed{K}}^{\rm ppE})$. Reference~\cite{Loutrel:2022xok} proposed ppE parameters for five harmonics, specifically
\begin{align}
    \ca_{0,0}^{\rm ppE} &= -q_{2} + 2q_{1} \cot\beta_{0} \sin\Delta_{a}\,,
    \\
    \ca_{+1,0}^{\rm ppE} &= 2q_{1} \left(2\cos\beta_{0}-1\right)\csc\beta_{0}\sin\Delta_{a} - q_{2}\tan\beta_{0}\,,
    \\
    \ca_{-1,0}^{\rm ppE} &= q_{1} \sin\Delta_{a} \left(2\cot\beta_{0} + \csc\beta_{0}\right) - q_{2} \left(1 + \frac{1}{2} \sec\beta_{0} \right)
    \\
    \ca_{+2,0}^{\rm ppE} &= \tan(\beta_{0}/2) \left(q_{2} \tan\beta_{0} - 2q_{1} \sin\Delta_{a}\right)\,,
    \\
    \ca_{-2,0}^{\rm ppE} &= \cot(\beta_{0}/2) \left(2q_{1}\sin\Delta_{a} - q_{2} \tan\beta_{0} \right)\,,
\end{align}
which all have exponent parameters $a_{\slashed{K}}^{\rm ppE} = 0$. Recall that $\Delta_{a} = a_{1} - a_{2}$, and $\beta_{0}$ is the initial inclination angle of the orbital angular momentum. The proposed ppE correction to the phase is given in Eq.~\eqref{eq:delta-Psi}, with
\begin{align}
    \cb &= \frac{5}{4} \sqrt{\frac{\pi}{5}} \chi_{Q} \left(q_{1} {\cal{U}}_{10} + q_{2} {\cal{U}}_{01} \right) \,,
    \\
    \label{eq:cc}
    \cc_{m',n} &= \frac{3\sqrt{5\pi}}{32} \chi_{Q} \Big[\left(n - m' - 2\Omega_{\epsilon}^{(0)}\right) \Gamma(q_{1},q_{2}) 
    \nn \\
    &- 2 \cos\beta_{0}\Delta \Omega_{\epsilon}\Big]\,,
\end{align}
where $({\cal{U}}_{10}, {\cal{U}}_{01})$ are complicated functions of $(\beta_{0},\omega_{0})$ given in Appendix~B of~\cite{Loutrel:2022ant}, and
\begin{align}
    \Gamma(q_{1},q_{2}) &= q_{1} \tan\beta_{0}\sin\Delta_{a} + \frac{1}{4} q_{2} \tan^{2}\beta_{0}\,,
    \\
    \Omega_{\epsilon}^{(0)} &= \frac{1}{4} \sec\beta_{0} \left[ 1 + 3\cos(2\beta_{0})\right]
    \\
    \Delta \Omega_{\epsilon} &= q_{1} \Delta\Omega_{\epsilon}^{(1)} + q_{2} \Omega_{\epsilon}^{(2)}\,,
    \\
    \label{eq:dOm1}
    \Delta\Omega_{\epsilon}^{(1)} &= -\frac{1}{4} \sec\beta_{0} \tan\beta_{0} \left[5 +3 \cos(2\beta_{0}) \right]\sin\Delta_{a}\,,
    \\
    \label{eq:dOm2}
    \Delta\Omega_{\epsilon}^{(2)} &= \frac{1}{32} \sec^{3}\beta_{0} \left[9 + 20\cos(2\beta_{0}) + 3\cos(4\beta_{0}) \right]\,,
\end{align}
Note that Eq.~\eqref{eq:cc} corrects a typo in Eq.~(120) of~\cite{Loutrel:2022xok}.

The full expression for the precession variables $(\beta, \alpha, \epsilon)$ are given in~\cite{Loutrel:2022ant}. In the weak asymmetry limit $q_{\cm} \ll 1$, these expression reduce to the form~\cite{Loutrel:2022xok}
\begin{align}
    \mu_{a} = \Omega_{a} \psi + \sum_{j=-2}^{2} C_{a}^{(j)}(q_{\cm}, \beta_{0}) e^{ij\psi} + {\cal{O}}(q_{\cm}^{2})\,,
\end{align}
where $\mu_{a} = [\sin\beta, \alpha, \epsilon]$. For completeness, we here provide the expression for the coefficients $C_{a}^{(j)}$.
\allowdisplaybreaks[4]
\begin{align}
    \Omega_{\beta} &= 0\,,
    \qquad
    \Omega_{\alpha} = -1\,,
    \\
    \Omega_{\epsilon} &= \Omega_{\epsilon}^{(0)} + \Delta \Omega_{\epsilon}\,,
    \\
    C_{\beta}^{(0)} &= \sin\beta_{0} - \frac{1}{2} q_{2} \sin\beta_{0} + q_{1} \cos\beta_{0} \sin\Delta_{a}\,,
    \\
    C_{\beta}^{(1)} &= \left[C_{\beta}^{(-1)}\right]^{\dagger} = -\frac{i}{2} q_{1} e^{-i\Delta_{a}}  \cos\beta_{0}\,,
    \\
    C_{\beta}^{(\pm2)} &= \frac{1}{4} q_{2} \sin\beta_{0}\,,
    \\
    C_{\alpha}^{(0)} &= -\frac{\pi}{2} - a_{2} + q_{1} \cos\beta_{0} \sin\Delta_{a}\,,
    \\
    C_{\alpha}^{(1)} &= \left[C_{\alpha}^{(-1)}\right]^{\dagger} = -\frac{i}{2} q_{1} e^{-i\Delta_{a}} \left(\cos\beta_{0} - i \tan\beta_{0}\right)\,,
    \\
    C_{\alpha}^{(2)} &= \left[C_{\alpha}^{(-2)}\right]^{\dagger} = \frac{i}{8} q_{2} \left(2 + \tan^{2}\beta_{0}\right)\,,
    \\
    C_{\epsilon}^{(0)} &= \epsilon_{0} + 2 q_{2} \sec\beta_{0}
    \nn \\
    &-\frac{1}{32} q_{1} \left[14 \cos\Delta_{a} + 5 \cos(\Delta_{a} - 4 \beta_{0}) + 4 \cos(\Delta_{a} - 2\beta_{0}) 
    \right.
    \nn \\
    &\left.
    + 4 \cos(\Delta_{a}+2\beta_{0}) + 5 \cos(\Delta_{a}+4\beta_{0})\right]\csc\beta_{0} \sec^{2}\beta_{0}
    \\
    C_{\epsilon}^{(\pm 1)} &= -q_{1} \sec\beta_{0}\,,
    \\
    C_{\epsilon}^{(2)} &= \left[C_{\epsilon}^{(-2)}\right]^{\dagger} = \frac{i}{4} q_{2} \sec\beta_{0}\,.
\end{align}

\bibliographystyle{apsrev4-1}
\bibliography{refs}
\end{document}